# TOWARD AN AGENT BASED DISTILLATION APPROACH FOR PROTESTING CROWD SIMULATION*

**Lam Thu BUI and Van Vien MAC**

*Abstract—* This paper investigates the problem of protesting crowd simulation. It considers CROCADILE, an agent based distillation system, for this purpose. A model of protesting crowd was determined and then a CROCADILE model of protesting crowd was engineered and demonstrated. We validated the model by using two scenarios where protesters are varied with different personalities. The results indicated that CROCADILE served well as the platform for protesting crowd modeling simulation.

## I. INTRODUCTION

Crowd control poses a great challenge for authorities. We are usually expecting a well-managed crowd; however, it also usually has a surprise factor. If a disruption happened to the crowd, disorder will appear and hence cause a mess and casualties accordingly. The research question is how to predict, manage and control crowd behaviors? A part of the solution is modeling and simulation [4, 6, 7].

Crowd simulation has been a significant research topic. There have been several methods proposed for modeling and simulating the crowd including System Dynamics, Particle Systems, Cellular Automata and Multi-Agent Systems as well as commercial software systems i.e *Massive, AI.implant, Age of Empires 3,* and *Crowd – MAGS*. Among these, multi agent systems have emerged as a promising technology and play as the foundation of crowd simulation.

The modeling and simulation methods can be classified into three types of crowd simulation: (1) *Flow-based Approach*, modeling a crowd as continuous flow of fluid [8, 9]; (2) *Entity-based Approach*, being suitable for medium-sized and homogeneous crowds. The movement of individuals is affected by some global/local laws that are introduced to represent various physical/ social/ psychological influences, such as flocking; and (3) *Agent-based Approach*, being for both medium and small sized heterogeneous crowds.

In this paper, we propose to investigate the CROCADILE system, a well-known multi-agent based system for battlefield simulation and apply it for protesting crowd simulation. Our research focuses on *how to design the crowd model and map it to CROCADILE*. We presented a case study where all aspects of protesting crowd were quantified and installed to CROCADILE. Through a case study adopted from [5], we demonstrated a protesting crowd model and showed that CROCADILE can be appropriately extended towards civilian crowd simulation.

The paper structure is organized as follow: the second section describes CROCADILE system. The third and fourth ones are for modeling and simulation of protesting crowds with CROCADILE. The paper is concluded in Section V.

*Lam Thu BUI is with Le Quy Don Technical University, 236 Hoang Quoc Viet, Ha Noi, Vietnam (e-mail: lam.bui07@gmail.com).

Van Vien MAC is with Institute of Information Technology, Ha Noi, Vietnam (e-mail: vienmvit@gmail.com).

## II. CROCADILE SYSTEM

We overview the CROCADILE system, which is used for modelling the protesting crowd in this paper. CROCADILE is basically an agent-based simulation system [1]. This simulation system concerns on emergent behaviour of the system; it represents the environment at an abstract level and therefore provides a broad perspective. The main focus of CROCADILE is that agents are working and interacting based on the behaviour weight system. The outcome of the simulations emerges from the interaction among the agents. That is why CROCADILE is called as an agent-based "*distillation*". Note that the *concept of "distillations"* is understood as any agent-based approach that concentrates on interaction between agents and does not concentrate as much on individual agents (see also [2, 3, 5]). CROCADILE applies quite simple techniques to control the motion of agents on the field, for example calculating the movement vector for agents based on the behaviour weight vectors. Under the control of these techniques, the motion of the agents can often be quite different from the reality of entities. It has been widely investigated in the area of military simulation where agents, representing soldiers, are hard-wired strict rules. There has been a question of how to extend these systems toward civilian crowd modelling and simulation? It is the main topic reported in the paper.

The agents in CROCADILE are provided with a wide range of capabilities, such as sensor, weapon, command, communication, and movement. In addition, the agents can belong to different teams, so they can be friendly, neutral, or hostile with each other. Further, they can communicate or command their subordinates to do specific tasks (missions). Agents have different behaviours and a number of triggers that help them deal with current, arbitrary events.

## III. CROWD SIMULATION WITH CROCADILE

### A. Protesting crowd characterization

We propose a model of the following characteristics for all agents and mapping to CROCADILE:

| | Characteristics | Description | CROCADILE's capacity |
|---|---|---|---|
| **Physical** | Size, location, speed, sensor, communication, strength | Physically describe the crowd | Number of agents, Position, speed, sensor, communications, health |
| | Lethality | Level of violence and property damage Likelihood of injuries and Deaths | Weapon, Armor Strength |
| **Social** | Organization | How organized is the group? | Red-teaming structure |
| | Leadership | How established is the leadership? | 1.Commander-subordinate structure 2.Behavior: Caring the |

| | Cohesiveness | Have members of the crowd bonded with each other? | 1.Command considerations (Formation) 2.Design to maintain spacing |
|---|---|---|---|
| **Psychological** | Mission | Is the crowd goal-oriented? | Mission assignment |
| | Emotion (panic/ angry) | Is the crowd emotionally intense? | Caring behaviors: 1.Caring the allied 2.Caring the friend 3.Caring the opponent |
| | Aggressiveness | How much aggression an agent has? | 1.Behavior: Caring the opponent 2.Behavior:Caring the mission 3.Capability considerations |

Note that our selection of CROCADILE is just because it is a complete system allowing customization of the agents and environment; no need to engineer agents from the scratch.

*B. Scenario formulation*

We followed a similar formulation in [5]. A scenario for protesting crowd simulation has two sides: the police force and the protesting crowd. For the police, they are assigned to protect the objective, keep the formation, and prevent the protesters penetrating to the area and avoid clashing with the protesters at the minimum. The police only use the weapon when they are under an attack or are being approached closely by the protesters. They have two types of behaviors: *default* and *being-hit*. Their weapon includes the short guns, tear gas, water cannon and plastic bullets.

For the protesters, they can be divided into several groups and have the leaders. The groups can be classified based on the aggressiveness level [11]. We considered the following types of protester groups:

**Passive group**: the one with tendency to avoid the police. They always follow the leaders and maintain the distance with the partners. If being hit or collided, they will change to the behavior of running away from the police and cannot maintain the distance with the partners. They have two types of behaviors: *default* and *being hit*. The passive group usually does not have the capability of the weapon.

**Moderate group**: They are moderate people. They are organized, follow the leaders, and maintain the distance. They will not change their behavior during the protesting.

**Aggressive group**: this is for aggressive people. They might not follow the leader's orders and also might separate from their crowd. Their action might include attacking the police and might be more aggressive if being hit. They might be equipped with lethal and non-lethal weapon.

## IV. A CASE STUDY

We setup two *teams* (meaning *sides*): the police and protesting; each team might have several *groups*. The values were adjusted reflecting relative relationship among entities.

*A. Behavioral settings:*

**- The police's default behavior:** The police agents will be initialized with this behavioral setting (values are ranged from 0 to 1). There are a lot of behavioral parameters in CROCADILE; here we described only the ones that we setup different values from the default value (zero). When the value is equal zero, it means that behavior is neutral and will not have effects on the movement of the agent.

Here we setup only two behaviors reflecting this default behavior: "*Caring the opponent*" and "*Caring the allied*" with the weight values of 0.2 and 0.8 respectively. Also the "*Likelihood of using movement ability*" is set at 0.1. Note that *weight* is different from the likelihood concept.

**- The police's being-hit behavior:** The police will change their behavior adapting to the new situation as being attacked. The likelihoods of *using movement ability* and *using firing ability* are increased to maximum (1.0). The "*Caring the allied*" is unchanged, however the behavior of "*Caring the opponent*" are changed with more details using sub-behavioral settings.

Note that the sub-behavior's value ranges from -1 to 1; the negative value means the sub-behavior will have the opposite effect. For example instead of getting close to the opponent, the agent will run away.

| Behavior's name | Sub-behavior's name | Value |
|---|---|---|
| Caring the opponent | Getting close to the opponent | 0.1 |
| | Getting close to the opponent in the weapon range | 0.1 |
| | Getting close to the opponent in the sensor range | 0.1 |
| | Getting close to the wounded opponent | 0.1 |
| | Getting close to the wounded opponent in the weapon range | 0.1 |
| | Getting close to the wounded opponent in the sensor range | 0.1 |

**- The passive protester's default behavior:** We use only three behaviors "*Caring the opponent*", "*Caring the allied*" and "*Caring the leader*" with the values of 0.05, 0.45, and 0.5 respectively. This means they tend to follow the leaders and avoid the police.

**- The passive protester's being-hit behavior:** This behavior setting is to be used once the agent is hit. In more details, the agent will run away from the police and does not care its leaders or partners. The changes are applied only to all sub-behaviors of "*Caring the opponent*" with the value of -0.5

**- The moderate protester's default behavior:** This type of protesters has only one behavioral setting with behaviors Caring the opponent, Caring the allied, Caring the leaders, Caring the terrain are set 0.15, 0.35, 0.35, and 0.15 respectively

**- The aggressive protester's default behavior:**

| Behavior's name | Value | Sub-behavior's name | value |
|---|---|---|---|
| Caring the opponent | 0,4 | | |
| Caring the allied | 0,3 | | |
| Caring the leaders | 0,3 | Keep the leaders in the sensor range | 0,5 |
| | | Keep the leaders in the communication range | 0,5 |

**- The aggressive protester's being-hit behavior:** This behavior will be triggered whenever the agents got hit. They will be more aggressive and lawless. The values of behaviors *Caring the opponent, Caring the allied,* and *Caring the leaders* are 0.5, 0.25, and 0.25 respectively.

### B. Other major capability settings

**Weapon:** The weapon capabilities are included Plastic Bullet, Water Cannon, Tear Gas, Short gun, Baton, and Stone where baton and stone are for the protesters while others are for the police.

**Health:** A police agent is setup as twice stronger than a protester (200 units in comparison to that of 100 units for a protester) reflecting that the police is more well-trained than the protesters

### C. Simulation Scenarios

Our scenarios are assumed in a flat urban terrain (i.e the park, or city square). A group of protesters gathered for a while and they are facing the police. To counter the protesters, the police setups several obstacles (05 fence objects) separating the protesting and the protected area. They will act if the protesters break the fence and enter the area. The numbers of police and protesters are as follows:

| Police | | Protesters | |
|---|---|---|---|
| *Items* | *Amount* | *Items* | *Amount* |
| With plastic bullets | 5 | Passive agents | 25 |
| With Tear Gas | 5 | Moderate agents with Stone | 20 |
| With Water Canon | 5 | Aggressive agents with stone | 20 |
| With Short gun | 5 | Moderate agents with Baton | 10 |
| | | Aggressive agents with Baton | 5 |
| | | Leader agents with Stone | 1 |

Since the protesting crowd is very much depending on the leaders, we propose to investigate in two cases being equivalent with two types of protesting leaders:

- **Case 1: Moderate leaders**

With this case, the protesting leaders were quite moderate in which they tend to avoid the police and pay the care to their movement and formation of the crowd.

| Behavior's name | Value |
|---|---|
| Caring the opponent | 0,45 |
| Caring the terrain | 0,55 |

- **Case 2: Aggressive leaders**

The leaders are different. They are more aggressive and are keen to confrontation with the police, meaning the sub-behaviors of "Caring the opponent" are increased from 0 to 0.1

### D. Results and discussion

Based on the results logged during the simulation run, we reported the summary as in the following table:

| | Criteria | Simulated results | |
|---|---|---|---|
| | | Case 1 | Case 2 |
| 1 | Achieved the goal? | Yes | No |
| 2 | Dead protesters | 0 | 0 |
| 3 | Wounded protesters | 0 | 22/106 |
| 4 | Dead Police | 0 | 0 |
| 5 | Wounded police | 0 | 8/20 |
| 5 | Destroyed obstacles | 5/12 | 8/12 |
| 7 | Protester Health Damage | 0/10600 | 220/10600 |
| 8 | Police Health Damage | 0/4000 | 20/4000 |

It is quite clear that in the first case, the police was able to contain the crowd. Although the protesters damaged several obstacles, they cannot penetrate further after the fence. Hence, the police successfully achieved their goal. However, the first case demonstrated quite peaceful scenario. There was no use of the weapon during the time of simulation.

In the second case, the situation is quite similar except the protester leaders where they were more aggressive. The dynamic of the scenario was changed accordingly in which the police failed to prevent the protesters. It again showed the importance of the leadership in the protesting crowd. In this case, 22 protesters were wounded in comparison to 8 police. There were lots of violent confrontation between the police and protesters. Note that both police and aggressive protesters are triggered when being hit for adapting to the new situations; during the simulation these triggers were all activated. For the protesters, once being hit, they become more aggressive, hence the confrontation is more serious.

In summary, the results supported the usability of CROCADILE for crowd modeling and simulation. The agents in CROCADILE are equipped with behaviors and triggers, which are capable of modeling all protesting crowd characteristics

### V. CONCLUSION

This paper overviewed research on protesting crowd and the usage of agent-based distillation for crowd simulation. For the protesting crowd modeling and simulation, we summarized its characteristics and classification. Based on this knowledge, we proposed to design a mapping of the protesting crowd to the model supported by CROCADILE, an agent based distillation system. To validate it, we carried out a case study on different scenarios. The results indicated the appropriateness of CROCADILE for crowd simulation.